\documentstyle[prb,aps,multicol,epsfig]{revtex}
\begin{document}
\draft
\widetext
\title{Influence of Electron-Phonon Interaction on\\
Spin Fluctuation Induced Superconductivity  }

\author{T. S. Nunner$^{1,}$\cite{PA}, J. Schmalian$^{2,1}$,  
               and K. H. Bennemann$^1$}
\address{$^1$ Institut f\"ur Theoretische Physik,
  Freie Universit\"at Berlin, Arnimallee 14, \\
       14195 Berlin, Germany \\
$^2$ Department of Physics, University of Illinois at 
Urbana-Champaign, Urbana,  IL 61801}
\date{ \today}
\maketitle 
\begin{abstract} 
\leftskip 54.8pt
\rightskip 54.8pt
We investigate the interplay of the electron-phonon
and the spin fluctuation interaction in the superconducting state
of YBa$_2$Cu$_3$O$_{7}$.
The spin fluctuations are described within the nearly antiferromagnetic
Fermi liquid theory and the phonons are treated 
using a shell model calculation of all  phonon branches.
The electron-phonon coupling is calculated using 
rigidly displaced ionic potentials screened by a background dielectric constant
$\varepsilon_\infty$ and by holes within the
CuO$_2$ planes. We get a 
superconducting state with $d_{x^2-y^2}$-symmetry whose origin  
is antiferromagnetic spin fluctuations.
The phononic contribution of {\em all} phonon modes of the system
 to the d-wave pairing interaction is
attractive. This gives a positive isotope
effect. The size of the isotope exponent depends strongly on the
relative strength of the electron-phonon and spin fluctuation
coupling. Due to the strong electronic correlations no phononic
induced superconducting state of s-wave character, is possible. 
\end{abstract}   
\begin{multicols}{2}
\narrowtext    

\section{Introduction}

The mechanism responsible for the superconducting state of high-T$_c$
superconductors is still discussed highly intensively.
Cooper-pairing induced by spin fluctuations~\cite{MBP91,MUT90,BSS87} 
seems to be likely. Based on the analysis 
of NMR experiments~\cite{MMP90}, it  offers an  explanation for transport
 and optical experiments~\cite{SP96} and a variety of anomalies
in underdoped cuprates~\cite{SPS98,CS98} 
and furthermore yields a d$_{x^2-y^2}$-wave pairing state with a transition 
temperature $T_c$ of correct order of magnitude.~\cite{MP94}

However, there are also observations, which indicate effects due to
electron-phonon interaction. Raman scattering experiments and IR
measurements show the softening of some phonon modes as the
temperature is decreased below $T_c$~\cite{Feile89,LTC94}. In YBa$_2$Cu$_3$O$_7$ the
largest shift was found for the $340
cm^{-1}$  ($42{\rm meV}$) Raman-active mode of $B_{1g}$-symmetry,
which involves the out-of-plane vibration of the oxygen atoms O(II)
and O(III) in the copper-oxygen planes~\cite{MRS87}. 
For some oxygen-related phonon 
modes the onset of the frequency softening was found to start even
50-60 K above $T_c$~\cite{LTC92}. The temperature dependence of this anomaly was 
found to be similar to that of the NMR relaxation rate $(T_1T)^{-1}$,
which, in the latter case, is due to an anomalous behavior  of the spin  excitations, and 
indicates a possible coupling between phonons and spin degrees of
freedom~\cite{LTC92}.
Especially interesting is the observation of the anomalous oxygen isotope
effect in the high-$T_c$ superconductors~\cite{Franck94}. At optimal
doping with largest $T_c$ only a very small oxygen isotope exponent
$\alpha=0-0.07$ is observed, whereas in the underdoped and overdoped systems the
isotope exponent can reach values close to the BCS-value
$\alpha=\frac{1}{2}$.

Due to these experimental findings it is of interest to investigate in
more detail the influence of electron-phonon interaction on the
spin fluctuation induced superconducting state.
In this context it was shown that oxygen buckling modes support a
$d_{x^2-y^2}$-pairing state, whereas oxygen breathing modes will
support an s-wave state~\cite{ND96,DT96,BS96}. Hence, an
analysis of the role of the electron-phonon interaction in
the cuprates has to take into account all vibrational modes.

In this paper we investigate the interplay of the spin fluctuation and
the electron-phonon interaction in the superconducting state
of YBa$_2$Cu$_3$O$_{7}$. We find that taking into account both
spin fluctuation and electron-phonon interaction gives a
superconducting state of d$_{x^2-y^2}$-symmetry, while    
electron-phonon interaction alone would yield an anisotropic
s-wave pairing with a much smaller critical temperature.
The $d_{x^2-y^2}$-superconducting state is caused by
antiferromagnetic spin fluctuations and the critical temperature $T_c$
is reduced by simoultaneous consideration of phonons. Nevertheless we
find a positive isotope effect. We show that the momentum dependence
of the total electron-phonon interaction amplifies the pairing
interaction by an amount of about 5-10\% in the $d_{x^2-y^2}$-wave
channel, which is a necessary condition for a positive isotope
effect. Furthermore we identify the phonon-modes, which make the most
dominant contribution to the $d_{x^2-y^2}$-pairing interaction. The
decrease of $T_c$ is explained by the large isotropic part of the
electron-phonon interaction, which enhances the electron scattering rate
significantly. 
However, this effect will be reduced when correlation effects are
taken into account, since correlation induced suppression of charge
fluctuations will predominantly reduce the isotropic  part of the
electron-phonon interaction.
Our  results are obtained with a self-consistent strong coupling
Eliashberg calculation.
Finally, in order to make this interplay of spin fluctuation and
electron-phonon interaction more transparent, we present a simple
weak coupling model.

In the present work we will not address a variety of rather important 
aspects of the electron-phonon interaction in cuprates.  
In calculating the electron-phonon coupling constants, we will completely
ignore the correlated nature of the involved charge carriers, even though
this is  expected to be essential   to deduce 
 reliable quantitative information about the strength of this  
  interaction and  its relative importance  for different phonon
modes. Based on calculations by Zeyher and Kuli\'c~\cite{ZK96},  we can at least show that
the latter aspect does not oppose to the conclusions of  our results.
Furthermore, lattice anharmonicities  will also be  neglected, since we believe that 
within the presented framework, where  the mutual influence of electron-
phonon and spin fluctuation interaction  is investigated within an Eliashberg theory,
anharmonicities, if relevant for the cuprates at all,  can  hardly
affect our result.
These rather drastic approximations are necessary to enable us to draw 
quantitative, material specific conclusions about the role of phonons 
within a spin fluctuation scenario.
The main complication in this context is that, in distinction to
the electronic system, the phononic degrees of freedom behave 
completely three dimensional, and there is no obvious restriction to
a few relevant phonon modes.

\section{Theory}

The coupled electron-phonon system is characterized by the  
Hamiltonian:
\begin{eqnarray}
H&=&
\sum_{{\bf k} \sigma} \varepsilon_{\bf k} 
c_{{\bf k}, \sigma}^\dagger c_{{\bf k}, \sigma}
+ g_{sf}  \sum_{{\bf k q }\sigma \sigma'}
c_{{\bf k}, \sigma}^\dagger \vec{\sigma}_{\sigma, \sigma'}
c_{{\bf k}+{\bf q}, \sigma'} \cdot \vec{S}_{-{\bf q}}
 \nonumber \\
& &  
+ \sum_{{\bf k,q} \lambda \sigma} g_{\lambda }({\bf q})
c_{{\bf k}, \sigma}^\dagger c_{{\bf k}+{\bf q}, \sigma}
\left(b_{{\bf q},  \lambda}^\dagger +b_{{\bf q}, \lambda } \right)
  \nonumber \\
& &+\sum_{{\bf q }\lambda}\omega_{{\bf q}, \lambda}  
b_{{\bf q},  \lambda}^\dagger b_{{\bf q}, \lambda }\enspace .
\label{Hamiltonian}
\end{eqnarray}
Here, $c_{{\bf k} \sigma}^\dagger$ is the electron creation operator
of the hybridized Cu-3d$_{x^2-y^2}$  and O-2p$_{x,y}$ orbitals
with energy $\varepsilon_{\bf k}$
describing the fermionic low energy degrees of freedom~\cite{ZR88}.
The spin fluctuations are  treated  in terms of the
interaction of the electron
spin with a spin-field $\vec{S}_{{\bf q}}$ characterized by a 
spin susceptibility $\chi({\bf q},\omega)$. Note, alternatively one could
also use a   Hubbard like Hamiltonian   and  determine $\chi({\bf q},\omega)$
diagrammatically, e.g. within the fluctuation exchange approximation~\cite{FLEX},
 which would not change
the conclusions of this paper.
The present method has the advantage that $\chi({\bf q},\omega)$
can be chosen  in agreement with  the experimentally observed spin susceptibility
of YBa$_2$Cu$_3$O$_{7}$.
In Eq.~(\ref{Hamiltonian}) $b_{{\bf q} \lambda}^\dagger$ is
the creation operator of a phonon in the vibrational branch $\lambda$ with energy 
$\omega_{{\bf q}, \lambda}$ and $g_{\lambda }({\bf q})$ is the 
electron-phonon coupling constant of the $\lambda$-th phonon branch
with the above Cu-O hybrid states.
For the electronic band dispersion we use
\begin{equation}
\epsilon({\bf{k}}) = -2 t  \left( \cos k_x  + \cos k_y \right) 
-4 t'  \cos k_x   \cos k_y   \;,
\end{equation}
 with the nearest neighbor hopping $t=250 \, {\rm meV}$ and
the next nearest neighbor hopping $t'=-0.45 t$, which reproduce
the bandwidth and Fermi surface shape of YBa$_2$Cu$_3$O$_{7}$.

The short ranged antiferromagnetic spin fluctuations
are described within the nearly antiferromagnetic fermi liquid 
 approach by the dynamical spin susceptibility~\cite{MMP90}:
\begin{equation}
{\rm Im} \chi({\bf q},\omega)={\rm Im} \left(
\frac{\chi_Q}{1+\xi^2({\bf Q}-{\bf q})^2
-i\omega / \omega_{\rm sf}  }\right)\, .
\label{MMP}
\end{equation}
Here, values for the antiferromagnetic correlation length $\xi=2.3a$,
the spin fluctuation energy $\omega_{\rm sf}=14meV$ 
and $\chi_Q=75meV$ are chosen as in Ref.~\cite{MP94}, which 
are deduced from  the analysis of NMR and NQR experiments
of YBa$_2$Cu$_3$O$_{7}$.
The resulting pairing interaction of the spin fluctuation 
interaction
is given by:
\begin{equation}
V^{sf}_{\bf q}(\omega)=g_{sf}^2 \chi({\bf q},\omega)\, .
\label{Vs}
\end{equation}
The   electron-spin fluctuation coupling constant 
$g_{sf}=0.68{\rm eV}$ for an electron concentration of $n_e=0.8$ is
chosen as in Ref.~\cite{MP94}, which was shown to yield $T_c=90K$ and
quantitative agreement with transport experiments, if one assumes that
these are entirely due to spin fluctuations~\cite{SP96}.

The phonons are described within a shell model, where the dynamical
matrix is obtained from the derivative of the lattice potential,
which consists of Coulomb potentials with core and shell charges,
Born-Mayer repulsive potentials and core-shell polarizabilities.
Details of the calculation are the same as in Ref.~\cite{WC66}
For the description of YBa$_2$Cu$_3$O$_{7}$ we use parameters 
for the core and shell charges, Born-Mayer parameters
and polarizabilities as given in Ref.~\cite{HLK93}.
The shell model was shown to give excellent agreement with
experimentally  determined phonon dispersion curves  for
several cuprate systems and is based on a rather small set of
physically well motivated parameters. 
The corresponding electron-phonon coupling is calculated using a model
recently proposed by Zeyher~\cite{Z90}, which uses rigidly displaced
ionic potentials screened by a dielectric constant $\epsilon_{\infty}$
and by holes within the $CuO_2$-planes. We have modified this model
to allow for the explicit consideration of local field corrections
taking the relative position of the screening $CuO_2$-planes and the
vibrating ions explicitly into account. The coupling of
the phonon-mode $\lambda$ to an electron localized at the site $\kappa$
within the primitive cell is given by:
\begin{eqnarray}
g_{\kappa \lambda}({\bf q})&=&-\sum_{\kappa' \alpha}
      \frac{Z_{\kappa'}e^2}{v_c^{2/3}} 
      \sqrt{\frac{\hbar}{2\omega_{\lambda}({\bf q})}} 
      \left(
      \frac{e_{\alpha \kappa' \lambda}({\bf q})}{\sqrt{M_{\kappa'}}} 
      \Phi^{(2)}_{\kappa \kappa' \alpha}({\bf q}) \right. \nonumber \\
& & \left.      - \frac{e_{\alpha \kappa \lambda}({\bf q})}{\sqrt{M_{\kappa}}} 
      \Phi^{(1)}_{\kappa \kappa' \alpha}(0)
      \right)     \, .
\end{eqnarray} 
$M_{\kappa}$ and $Z_{\kappa}$ are the mass and the charge of the ion
with basis index $\kappa$, $v_c$ is the volume of the primitive cell, 
$\omega_{\lambda}({\bf q})$ and $e_{\alpha \kappa \lambda}$ are the
vibration frequency and the normalized eigenvector of the
phonon mode $\lambda$ with wave vector ${\bf q}$ and $\alpha$ is 
a Cartesian index. $\Phi^{(1(2))}_{\kappa \kappa' \alpha}$ denotes
the contribution of the of the $\kappa'$-ion to the total Coulomb
potential at lattice site $\kappa$, which is caused by the
vibration of the ion $\kappa'$ and of the ion $\kappa$, respectively:
\begin{eqnarray}
\Phi^{(1(2))}_{\kappa \kappa' \alpha}({\bf q})&=& \frac{i}{v_c^{1/3}}
       \sum_{\bf G_1 G_2} \epsilon^{-1}({\bf k+G_1,k+G_2})
\nonumber \\ & &
       v({\bf k+G_2})({\bf k+G_{1(2)}})_{\alpha}
       e^{i{\bf G_1}\kappa}e^{-i{\bf G_2}\kappa'}   \, .
\end{eqnarray}
${\bf G_1,G_2}$ are reciprocal lattice vectors and
$v({\bf k})=\frac{e^2}{\epsilon_0 {\bf k}^2}$ denotes the bare Coulomb
interaction. The inverse dielectric function $\epsilon^{-1}$ will be
discussed in the appendix.

For simplicity we assume, that the Cooper pair forming electrons
are mainly located at the in-plane Cu-sites. Therefore only the coupling
$g_{\kappa \lambda}$ where $\kappa$ refers to the Cu(II)-site will be
considered. This index $\kappa$ can be suppressed in the following. 
The electron-phonon induced pairing interaction is given by:
\begin{equation}
V^{ph}_{\bf q}(i \omega_n) = \frac{1}{N_z}
\sum_{q_z , \lambda} g_{\lambda }({\bf q}, q_z)^2 
\frac{2 \omega_{{\bf q}, q_z, \lambda}}
{(i \omega_n)^2 - \omega_{{\bf q}, q_z, \lambda}^2} \, .
\label{eq:v_ph}
\end{equation}
Here, the summation with respect to $q_z$ takes into account that
the pairing interaction 
can also be generated by phonons   propagating perpendicular 
to the CuO$_2$ planes, even if 
the paired electrons  are within these planes.

The total electron-phonon coupling strength may then be characterized e.g. via
the dimensionless parameter
\begin{equation}
\lambda=2 \int_{0}^{\infty} \frac{\alpha^2F(\omega)}{\omega} d\omega
             \enspace ,
\label{eq:lambda}
\end{equation}
where the McMillan function $\alpha^2F$ is defined by:
\begin{eqnarray}
\alpha^2F(\omega)&=&\frac{1}{\sum_{\bf k}\delta(\epsilon_{\bf k})}
 \sum_{\bf kk'} 
  \delta(\epsilon_{\bf k})\delta(\epsilon_{\bf k'}) \nonumber \\ & &
  \sum_{\lambda} |g({\bf k,k'},\lambda)|^2
  \delta(\omega-\omega_{\lambda}({\bf k-k'}))\, .
  \end{eqnarray}

Having discussed the spin fluctuation and electron-phonon induced 
interactions, we investigate the superconducting state within the
strong coupling Eliashberg theory, where as usual, the self energy
is expanded  with respect to unitary
and Pauli matrices in Nambu space ${\hat \tau^i}$:
\begin{equation}
{\hat \Sigma}_{{\bf k} }(  i\omega_n) =  Y_{{\bf k}}( i\omega_n)
{\hat \tau^0}+ X_{{\bf k}}( i\omega_n) {\hat \tau^3} +
\Phi_{{\bf k} }( i\omega_n){\hat \tau^1} \; .
\end{equation}
$Y_{{\bf k}  }(i\omega_n)=i\omega_n 
\left(  1-Z_{{\bf k} }( i\omega_n) \right)$ and $X_{{\bf k} }(i\omega_n)$ refer to the
self energy renormalizations which occur also in the normal state and 
$\Phi_{{\bf k}}(i\omega_n)=
\Delta_{{\bf k}}(i\omega_n)Z_{{\bf k}  }(i\omega_n)$ to the 
anomalous self energy which is nonzero only in the superconducting state.
$\omega_n=(2n+1) \pi/\beta$ are the fermionic Matsubara frequencies, with
$\beta=1/T$. The self energies  are 
determined by   the  Eliashberg equations:
\begin{eqnarray}
\Phi_{{\bf k}}(i\omega_n) &=&\frac{1}{\beta} 
\sum_{{\bf k}',n'}
\frac{\left( {\tilde V}_{  {\bf k},{\bf k}'}(i\omega_n-i\omega_{n'})  
  \right) 
\Phi_{{\bf k}' }(i\omega_{n'})} {D_{{\bf k}'  }(i\omega_{n'})} \, , 
\nonumber  \\
X_{{\bf k} }(i\omega_n) &=&\frac{1}{\beta} \sum_{{\bf k}' ,n'}
\frac{V^{}_{  {\bf k},{\bf k}'}(i\omega_n-i\omega_{n'})   
(\varepsilon_{{\bf k}'  } +X_{{\bf k}' }(i\omega_{n'}))} {D_{{\bf k}' , \lambda'}(i\omega_{n'})} \, , 
\nonumber  \\
Y_{{\bf k} }(i\omega_n)  &=&\frac{1}{\beta} 
\sum_{{\bf k}' ,n'}
\frac{V^{}_{  {\bf k},{\bf k}'}(i\omega_n-i\omega_{n'})  
i\omega_{n'}Z_{{\bf k}' }(i\omega_{n'})} {D_{{\bf k}'  }(i\omega_{n'})} \, , 
\label{eq:Eliashberg}
\end{eqnarray}
where 
\begin{equation}
{\tilde V}_{ {\bf k},{\bf k}'   }(i\omega_n )=V^{sf}_{  {\bf k}-{\bf k}'  }(i\omega_n )+
V^{ph}_{  {\bf k}-{\bf k}'  }(i\omega_n )\, ,
\end{equation}
\begin{equation}
  V_{  {\bf k},{\bf k}'  }(i\omega_n )=V^{sf}_{  {\bf k}-{\bf k}'   }(i\omega_n )-
V^{ph}_{  {\bf k}-{\bf k}'  }(i\omega_n )\, ,
\end{equation}
and the denominator is given by
\begin{eqnarray}
D_{{\bf k}'  }(i\omega_{n'})&=&
(i\omega_{n'} Z_{{\bf k}' }(i\omega_{n'}))^2\nonumber \\ & &-
(\varepsilon_{{\bf k}' }
+X_{{\bf k}' }(i\omega_{n'}))^2 -
\Phi_{{\bf k}' }(i\omega_{n'})^2 .
 \end{eqnarray}
After analytical continuation to the real frequency axis,
$i \omega_n=i(2n+1)\pi/\beta \rightarrow \omega +i0^+$,
this set of coupled equations is solved self consistently
using the numerical framework of Ref.~\cite{SLG96}.
We do not restrict the momentum summation to the
Fermi surface but take the entire  BZ into account.
Since  both pairing interactions $V^{sf}$ and $V^{ph}$ are chosen to
reproduce the experimental spin-susceptibility and phonon-dispersion,
we do not take additional renormalizations of $V_{  {\bf q} }(i\omega_n )$ and
${\tilde V}_{  {\bf q} }(i\omega_n )$ into account.

\section{Results}

\begin{figure}
\centerline{\epsfig{file=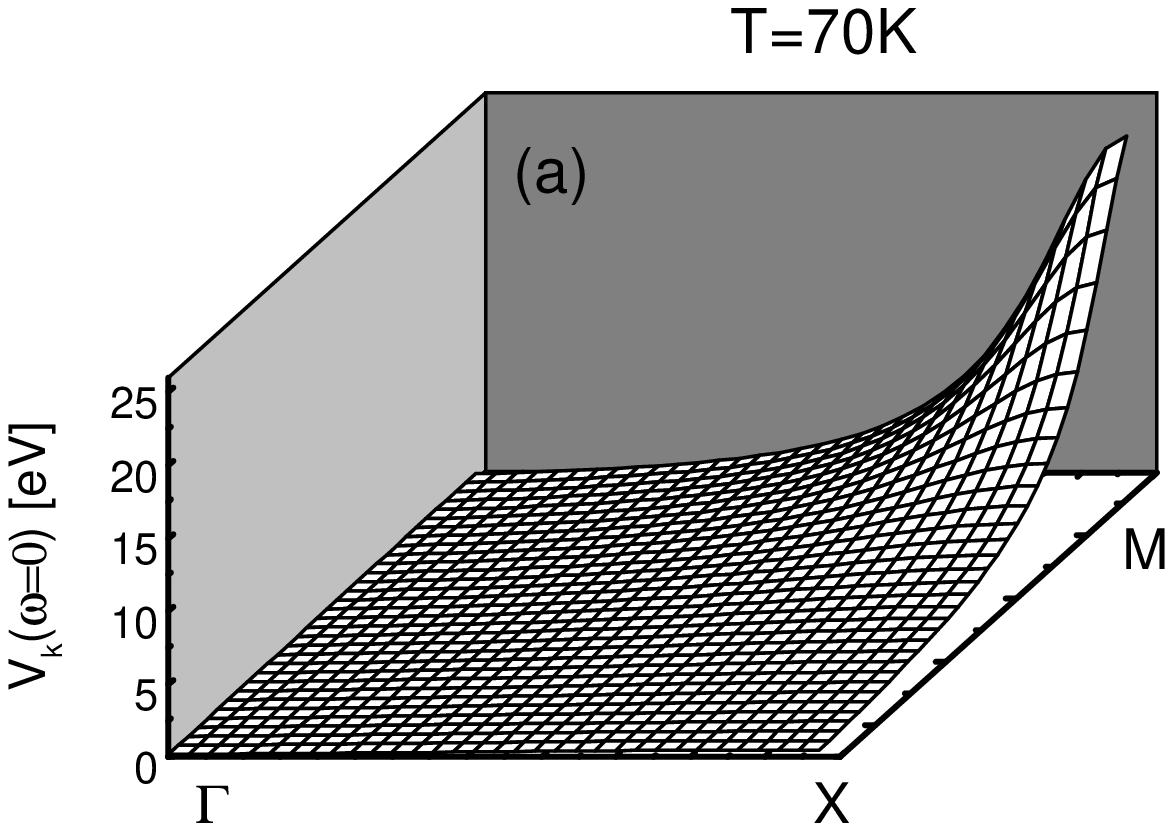,width=8cm,bb=110 480 500 740}}
\centerline{\epsfig{file=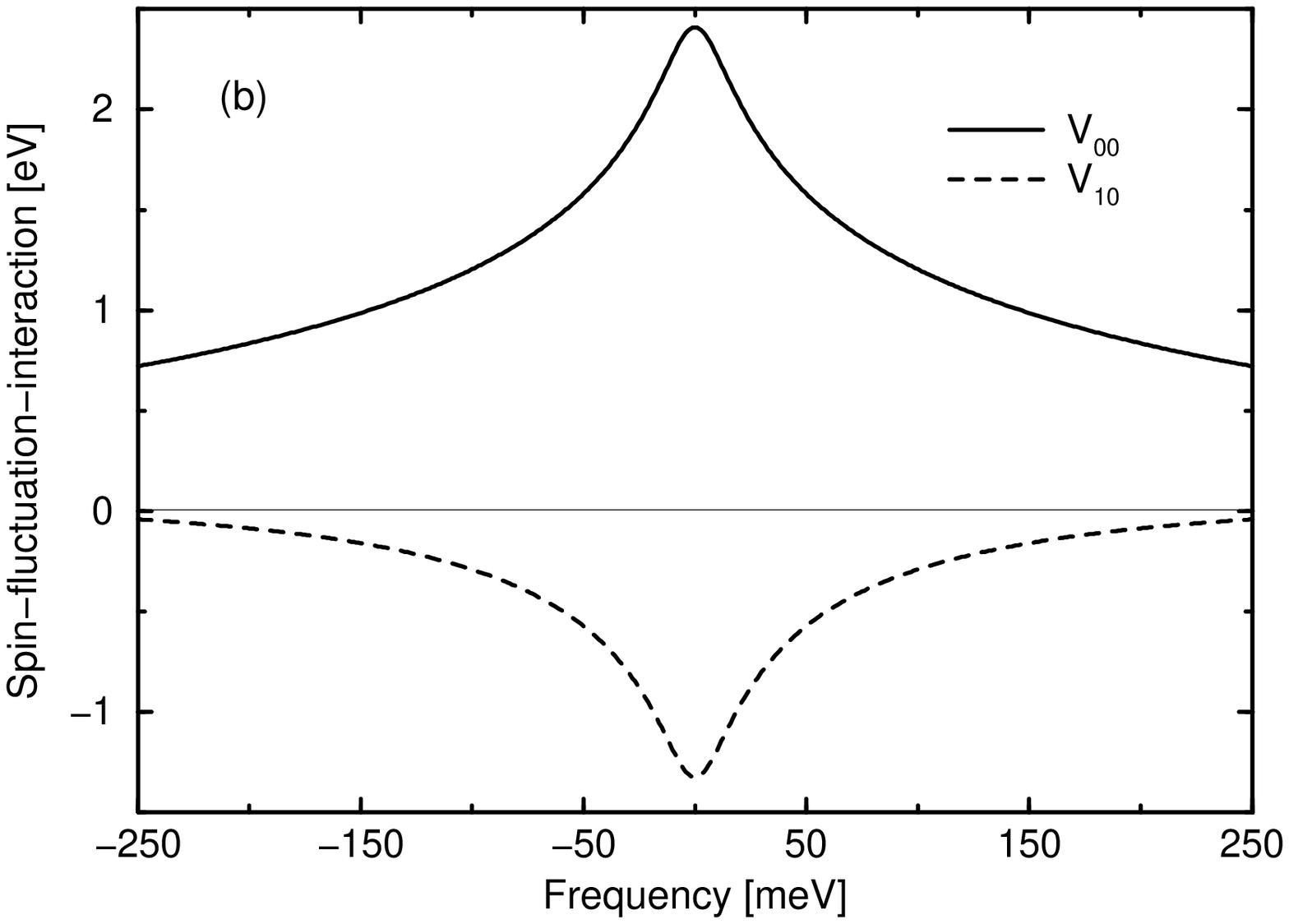,width=7cm,bb=40 50 540 440}}
\caption{Spin fluctuation interaction at T=70K: (a) momentum dependence of the
  spin fluctuation interaction for frequency $\omega=0$, (b) frequency
  dependence of the on-site interaction $V_{00}$ and the nearest
  neighbor interaction $V_{10}$.}
\label{fig_1}
\end{figure}

\begin{figure}
\centerline{\epsfig{file=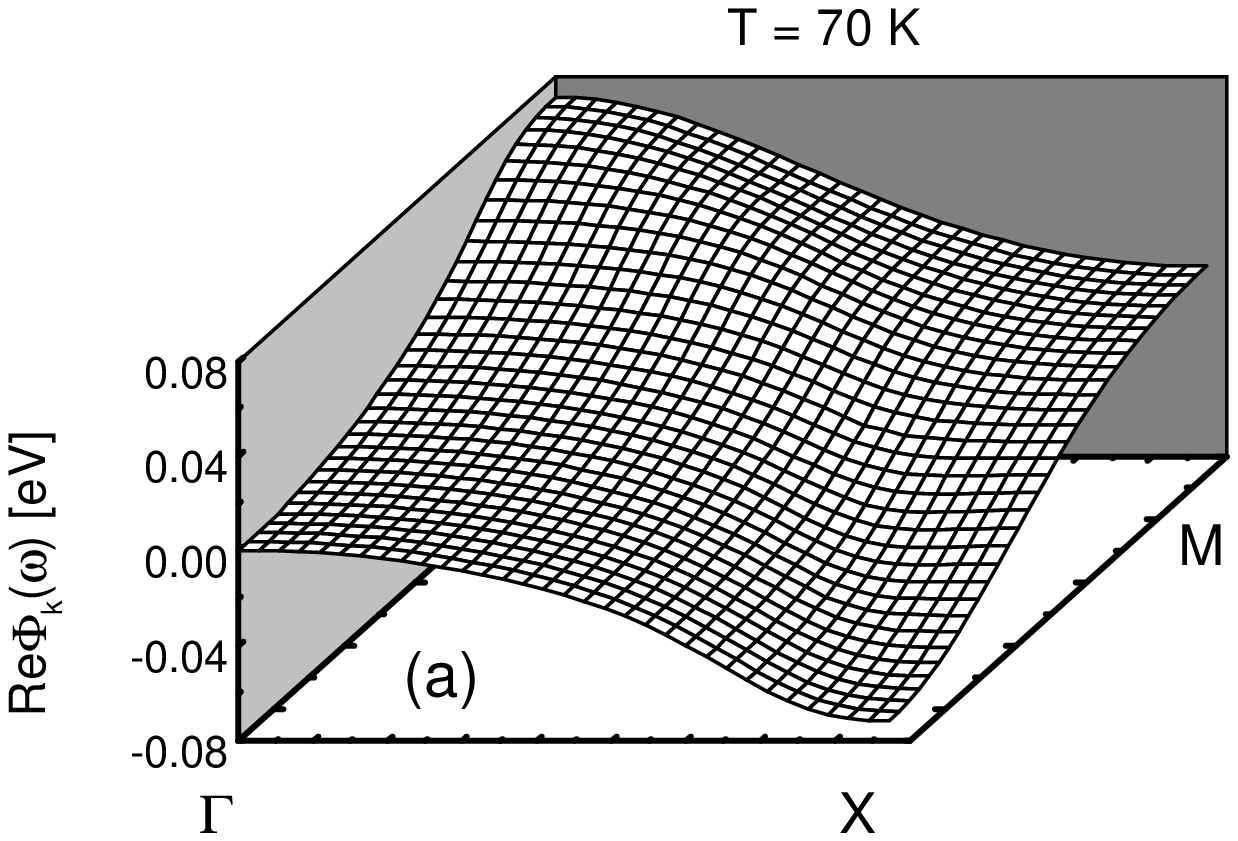,width=8cm,bb=110 490 500 750}}
\centerline{\epsfig{file=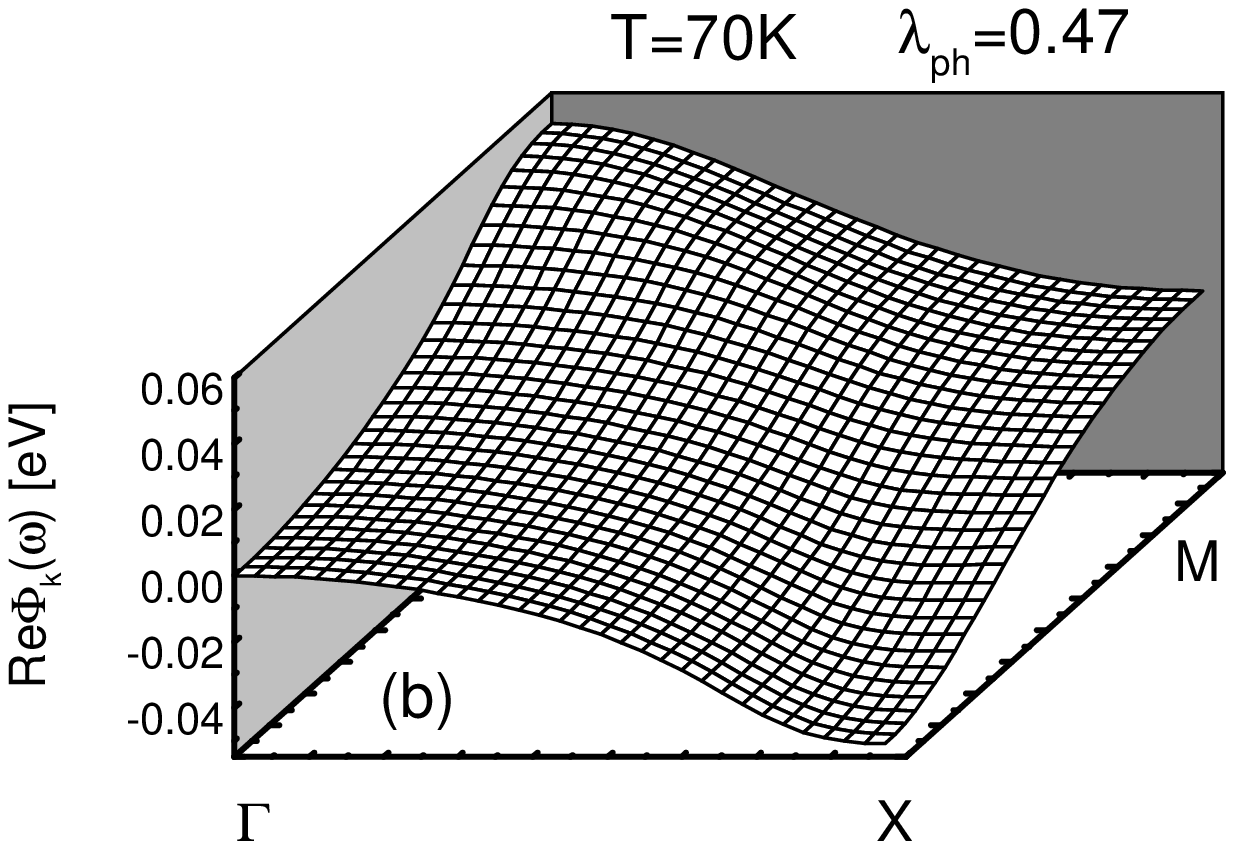,width=8cm,bb=110 465 500 740}}
\caption{Off diagonal self-energy $\Phi_k(\omega)$ for (a) only spin fluc\-tua\-tion 
  in\-duced super\-con\-ducti\-vi\-ty and (b) both phonon and
  spin fluc\-tua\-tion in\-duced
  super\-con\-duc\-ti\-vi\-ty. In both cases the super\-con\-duc\-ting state is
  characterized by $d_{x^2-y^2}$-symmetry order parameter, since 
   $\Phi_k \sim (\cos k_x - \cos k_y)$. Note, $\Phi_k(\omega)$
  is reduced due to the electron-phonon interaction.}
\label{fig_2}
\end{figure}

First, we will demonstrate that our treatment of the spin fluctuations yields indeed
a superconducting state with $d_{x^2-y^2}$  symmetry and a
  reasonable critical temperature.
The effective interaction due to spin fluctuations is repulsive for all momenta and
is peaked at the antiferromagnetic wave vector ${\bf Q}=(\pi/a,\pi/a)$, as
can be seen in Fig.~\ref{fig_1}. In
agreement with  Ref.~\cite{MP94}, the self-consistent solution of the
Eliashberg equations yields $d_{x^2-y^2}$-superconductivity, as can be
deduced from the momentum dependence of the
off-diagonal self-energy in Fig.~\ref{fig_2}(a), which exhibits the typical
form: $\Phi \sim (\cos k_x - \cos k_y)$.

Variation of the temperature shows that the superconducting
gap vanishes around 110K. Although this is slightly larger than the
experimental critical temperature of $YBa_2Cu_3O_7$,
we decided to accept this still reasonable value for $T_c$, since the
spin fluctuation interaction has already been chosen in agreement
with transport experiments~\cite{SP96} and manipulation of the
electron-spin fluctuation coupling constants would introduce new and
arbitrary parameters. 

In the following we investigate the influence of the electron-phonon
interaction on the spin-fluctuation induced superconducting state. 
Solving the Eliashberg equations yields
$d_{x^2-y^2}$-superconductivity even in the presence of phonons,
as can be seen from the momentum dependence of the gap-function in
Fig.~\ref{fig_2}. For obtaining some information about the
isotope effect, we have performed Eliashberg calculations for two 
different oxygen-  

\begin{figure}
\centerline{\epsfig{file=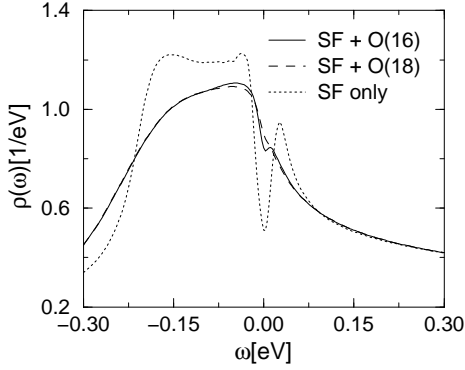,width=6cm,bb=30 10 280 250}}
\caption{Momentum avaraged electronic density of states at T=80K for a superconducting
  state induced only by spin fluctuations and by spin fluctuation and
  electron-phonon interaction together for two different
  different oxygen isotopes O(16) and O(18).} 
\label{fig_3}
\vspace{-.05cm}
\end{figure}

\noindent
isotopes O(16) and O(18). For this the whole electron-phonon
interaction, i.e. the phonon dispersions and the electron-phonon
coupling constants, has been calculated
 separately for the two oxygen
masses. In Fig.~\ref{fig_3} the momentum averaged density of states
at a temperature of $T=80K$
is shown for the two oxygen isotopes O(16) and O(18). For comparison
the density of states for only
spin fluctuation induced superconductivity is shown as well.
Two important things can be learned from Fig.~\ref{fig_3}: the
superconducting gap is reduced dramatically in the presence of
phonons, but nevertheless the isotope effect is positive. This can be
infered from the fact that there still remains a small
superconducting gap in the density of states of the oxygen isotope
O(16) at the temperature chosen in Fig.~\ref{fig_3} (T=80K),
whereas for O(18) the superconducting gap has already completely
disappeared. 

Note that similar results have been obtained by H.-B. Sch\"uttler and
C.-H. Pao~\cite{PaoSch}. They also find that the electron-phonon
interaction reduces the critical temperature of spin
fluctuation-induced superconductivity considerably, whereas the
isotope exponent remains small but negative. However in their
calculation this is due to the local nature of the electron-phonon interaction
whereas the focus of our work is to investigate the role of the
k-dependence of this interaction.

In order to obtain an estimate for the critical temperature of
the two oxygen isotopes, we have performed succesive calculations in
the temperature range between 78K and 83K. Since the self-consistent
solution of the Eliashberg equations gives rise to a finite single
particle scattering rate in the gap once the temperature is nonzero we choose
the disappearance of the dip structure in the DOS as criterion for
$T_c$. In this way we obtain for the oxygen isotope O(16) a critical
temperature of $T_c=82K$ and for the isotope O(18) $T_c=80K$. 
Using these values for the calculaion of the isotope exponent
$\alpha=- \frac{\partial (\ln T_c)}{\partial (\ln m)} \approx -
\frac{\Delta T_c}{T_c} \frac{m}{\Delta m}$, we obtain an approximate
value of $\alpha \approx 0.2$,
which is only slightly larger than experimentally found in
optimally doped cuprates. 

\begin{figure}
\centerline{\epsfig{file=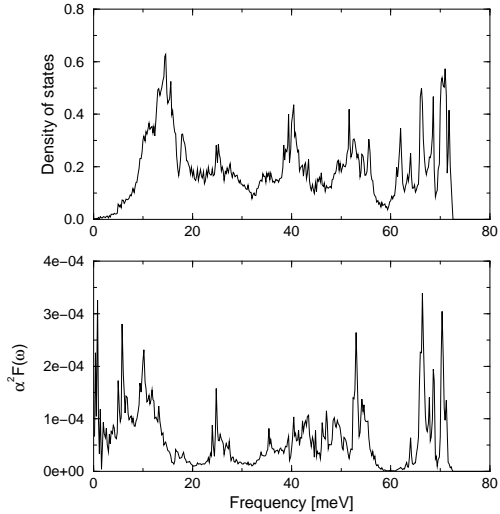,width=9cm}}
 \caption{ Phonon density of states for $YBa_2Cu_3O_7$
  obtained from a shell-model calculation using parameters as in
  Ref. [22],
 and the McMillan function $\alpha^2F(\omega)$.
  The calculation of the DOS and $\alpha^2F(\omega)$ were performed
  with a channel width of 0.2 meV.}
\label{fig_4}
\end{figure}

For a better understanding of the phononic contribution to the pairing
interaction and the reason for the positive isotope effect, the phonons will
be analyzed separately now. 
The phonon density-of-states and the McMillan function $\alpha^2F$ are
shown in Fig.~\ref{fig_4}.  
For total electron-phonon coupling strength we obtain
$\lambda=0.47$, which is in good agreement with estimates of
$\lambda_{tot}=0.4-0.6$ \cite{LTC91,FTC90} from
experimental determination of frequency shifts
and  line widths. Nevertheless, the calculated $\lambda=0.47$  should

\noindent
lead to a considerable contribution to the resistivity, which within
 the spin fluctuation model is believed to be dominated by purely
 electronic scattering. This indicates that role and relative weight
 of these two scattering channels for transport phenomena is not yet completely 
understood. We checked that modifications of  $\lambda$  due to
backscattering vertex-corrections do not reduce $\lambda$  considerably.

\begin{figure}[p]
\centerline{\epsfig{file=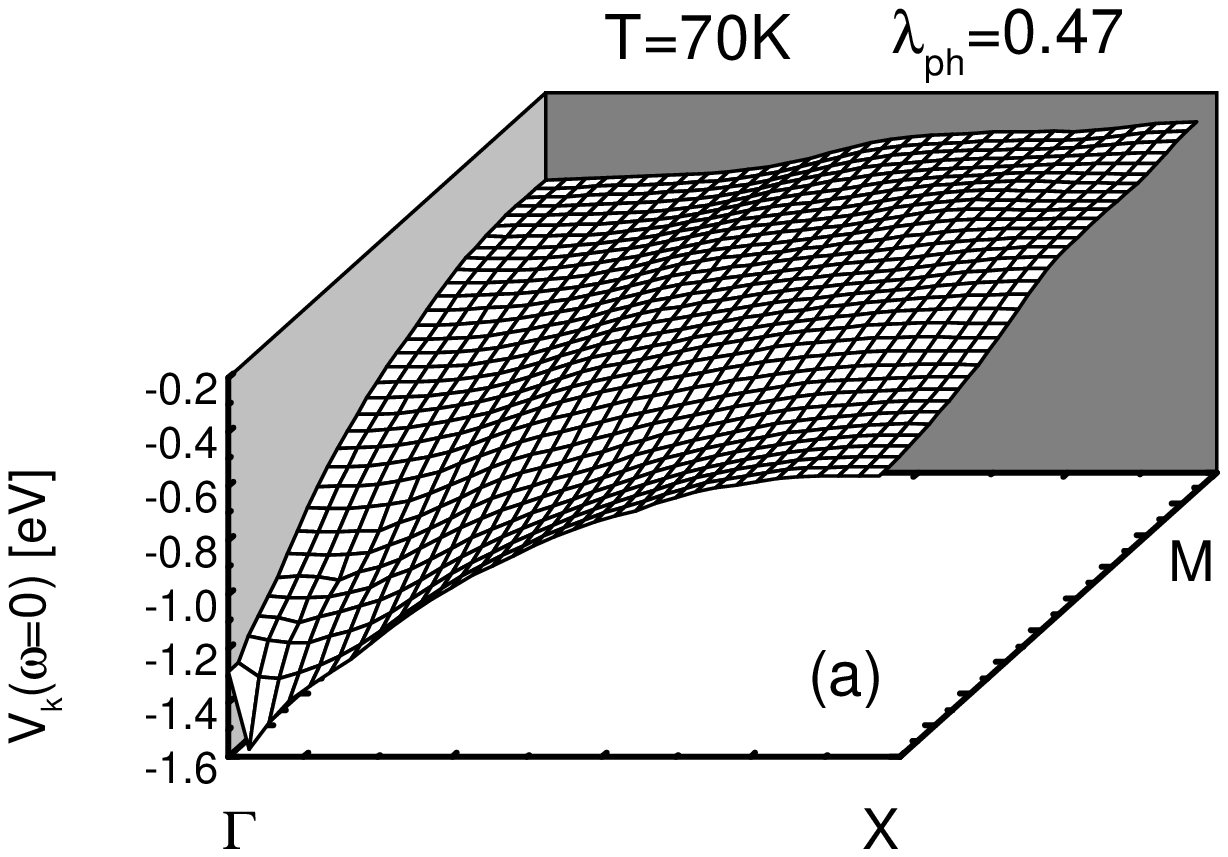,width=8cm,bb=90 460 500 740}}
\centerline{\epsfig{file=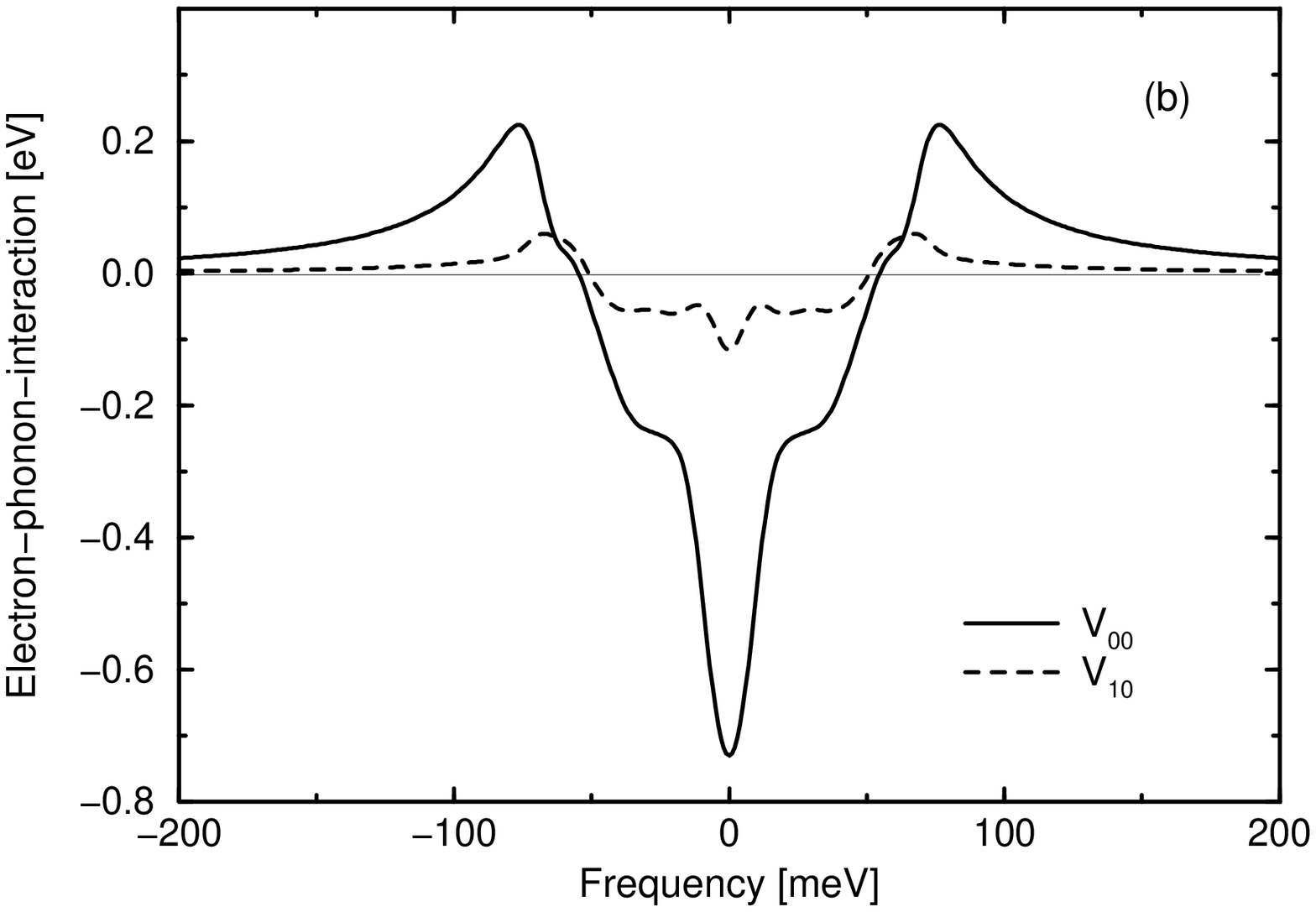,width=7cm,bb=40 70 540 440}}
\caption{Electron-phonon interaction at T=70K and $\lambda_{ph}=0.47$:
  (a) momentum dependence of the
  electron-phonon interaction $V_k$ for frequency $\omega=0$, (b) frequency
  dependence of the on-site pairing interaction $V_{00}$ and the nearest
  neighbour pairing interaction $V_{10}$.}
\label{fig_5}
\end{figure}

In Fig.~\ref{fig_5}(a), we show  the momentum dependence of the zero frequency
 effective phonon mediated
electron-electron interaction of Eq.~(\ref{eq:v_ph}) . The
interaction is attractive for all momenta and most important for the
present study peaked for small
momentum transfer. In Fig.~\ref{fig_5}(b)

\noindent
 we show the
on site interaction $V_{00}$:
\begin{equation}
V_{00}=\frac{1}{N} \sum_{k_x,k_y} V(k_x,k_y) \enspace,
\end{equation}
which represents the pairing interaction in the case of
isotropic s-wave superconductivity, and the nearest
neighbor interaction $V_{10}$:
\begin{equation}
V_{10}=\frac{1}{N} \sum_{k_x,k_y} V(k_x,k_y) \cos k_x \enspace, 
\label{eq:V10}
\end{equation}
which represents the pairing interaction in the case of $d_{x^2-y^2}$-symmetry
superconductivity~\cite{note1}. Both $V_{00}$ and $V_{10}$ are
attractive in some frequency range, but the on site interaction is
much stronger and the corresponding attractive frequency range is
slightly larger. Therefore, phonons alone will give rise to s-wave
superconductivity. This is natural for a phononic mechanism, since
the electron-phonon interaction is attractive for all momenta and 
thus the momentum averaged interaction  $V_{00}$  is dominant.
The self-consistent solution of the Eliashberg equations confirms
this. 
Taking only the electron-phonon interaction into account, we  
obtain  superconductivity with anisotropic s-wave
symmetry without nodes 

\begin{figure}
\centerline{\epsfig{file=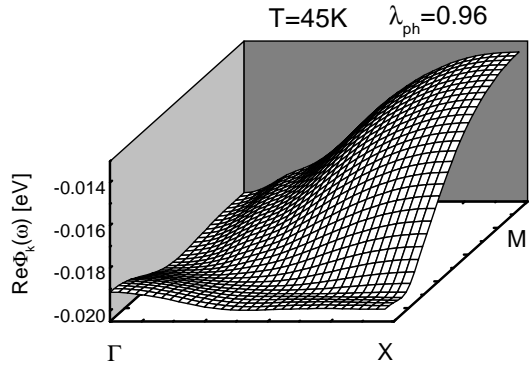,width=8cm,bb=90 460 500 740}}
\caption{Off diagonal self-energy for only phonon induced
  superconductivity. The momentum dependence exhibits clearly an
  anisotropic s-wave symmetry.}
\label{fig_6}
\end{figure}

\noindent
as can be  seen from the momentum dependence of
the off-diagonal self energy in Fig.~\ref{fig_6}. For obtaining
this result
at the numerically accessible temperatures $T>30\, {\rm K}$
 we had to double the total electron-phonon coupling 
strength artificially. But we expect that enhancing the 
coupling strength should solely increase the critical temperature 
and  not change the symmetry of the superconducting
state. Even for this artificially large coupling strength of $\lambda=0.96$
superconductivity already vanishes between about $T=45-50K$. Hence,
for realistic coupling strengths the critical temperature of phonon
induced superconductivity is much smaller than the critical temperature of
spin fluctuation induced superconductivity. Note that in this case
the repulsive Coulomb interaction was not taken into account, which
will suppress the critical temperature of the s-wave superconducting
state even more.

Most interesting, however, is the result that all phonon modes
together can generate an attractive interaction between nearest
neighbor sites. Thus the consideration of electron-phonon coupling
increases the pairing interaction in the $d_{x^2-y^2}$-wave channel,
which is a necessary prerequisit for a positive isotope effect.  
The reason for the resulting decrease of the superconducting gap due to
the presence of phonons is the large on-site interaction $V_{00}$,
which enhances the scattering rate dramatically. Therefore, concerning the sign
of the isotope effect, it is plausible, that the isotope effect
is positive as long as the positive contribution from the increased 
pairing interaction $V_{10}$ (where the larger cut-off
corresponds to the smaller oxygen mass) outweighs
the negative contribution from the increased scattering rate due to
larger on-site interaction $V_{00}$.
Also it is clear that including both spin fluctuation and
electron-phonon interaction still yields
$d_{x^2-y^2}$-superconductivity, since the consideration of
electron-phonon coupling even increases the $d_{x^2-y^2}$-pairing
interaction, whereas the spin fluctuation contribution to $V_{00}$,
the s-wave pairing interaction, is strongly repulsive, as can be seen
from Fig.~\ref{fig_1}, and therefore reduces the s-wave pairing
interaction dramatically. 

We will focus now on the question, why the nearest neighbor
electron-phonon interaction is attractive at all.
The attractiveness of $V_{10}$ can be
traced back to the fact that the phononic interaction is peaked for small
momentum transfer, since multiplication with $\cos k_x$ in
Eq.~(\ref{eq:V10}) gives more weight to small momenta.
Obviously, besides the shape of the Fermi surface, which is taken
from ARPES experiments, the momentum dependence of the electron-phonon interaction
is the key feature, which controls possible contributions to
different pairing symmetries. The momentum dependence of the
electron-phonon interaction consists of two contributions, as
can be seen by inspection of Eq.~(\ref{eq:v_ph}); namely 
the momentum dependence of the electron-phonon coupling constants and
the momentum dependence of the phonon-Green's function. Because the momentum
dependence of the phonon-Green's function is directly related to the
dispersion of the phonon spectrum, which is relatively weak in the case
of dominating optical phonon modes, we find that the momentum dependence of the
electron-phonon coupling constants is the decisive factor. 
Therefore, phonon modes with coupling strength largest for
small momentum transfer can enhance the $d_{x^2-y^2}$ pairing
interaction. However, phonon modes with coupling strongest for large momentum
transfer reduce the $d_{x^2-y^2}$ pairing interaction.   

\begin{figure}
\centerline{\epsfig{file=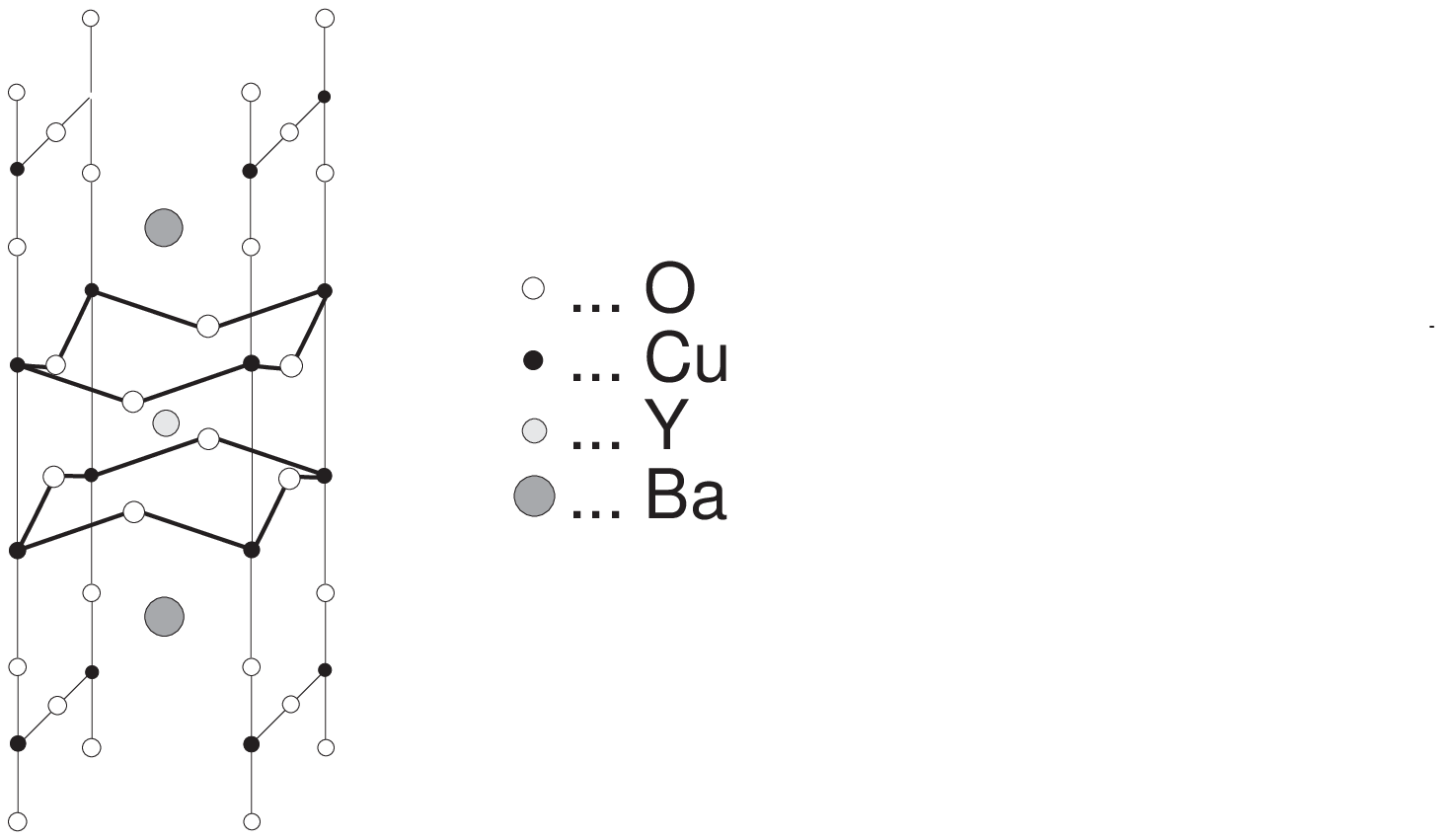,width=6cm,bb=100 460 330 710}}
\caption{Structure of $YBa_2Cu_3O_7$.}
\label{fig_7}
\end{figure}

\begin{figure}
\centerline{\epsfig{file=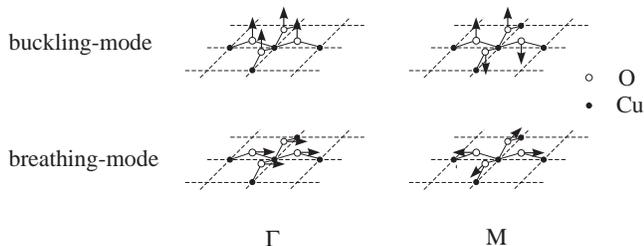,width=8.5cm,bb=56 530 380 680}}
\caption{Displacement patterns of the buckling mode and of the
  breathing mode at the $\Gamma$-point and the M-point of the BZ.}
\label{fig_8}
\end{figure}

\begin{figure}
\centerline{\epsfig{file=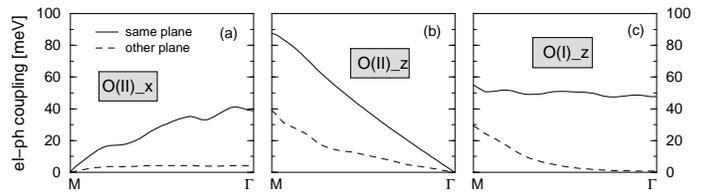,width=9.5cm,bb=0 50 580 230}}  
\caption{Electron-phonon coupling constant for the vibration of the
  in-plane oxygen O(II) (a) in x-direction, (b) in z-direction and
  (c) for the vibration of the apex-oxygen O(I) in z-direction. Since
  every hole is located just in one of the planes, the coupling is
  different for the vibrations of the oxygens of the two planes. The
  coupling of the oxygen in the same plane is denoted by a solid line
  and the coupling of the oxygen of the other plane by a dashed line. 
  Note, in the case of the apex-oxygen, this refers to the closer
  and the further apex-oxygen.}
\label{fig_9}
\vspace{-.5cm}
\end{figure}

In order to find out which phonon modes support the
$d_{x^2-y^2}$-pairing interaction, the vibrations of single atoms (per unit
cell) will be investigated. For illustration of the structure of
YBa$_2$Cu$_3$O$_7$ we refer to Fig.~\ref{fig_7}.
The greatest coupling strength is caused
by the vibration of the in-plane oxygens O(II), with vibration
amplitudes along the Cu-O bonding axis
and along the z-axis, i.e. breathing mode like and buckling
mode like vibrations, and by the vibrations of the apex oxygen O(I) in
z-direction. For illustration of the displacement patterns of the
breathing mode and the buckling mode we refer to
Fig.~\ref{fig_8}. 
In Fig.~\ref{fig_9} the momentum dependence along the 
$\Gamma \rightarrow M$-direction of the corresponding electron-phonon
coupling is shown.

\noindent
The in-plane oxygen O(II,III) vibrations are quite
well described by the momentum dependence $g({\bf k}) \sim (\sin k_x +
\sin k_y)$ and $g({\bf k}) \sim (\cos k_x + \cos k_y)$, which is the
result for the breathing and buckling mode, respectively, if only
nearest neighbor interaction is considered. Thus the breathing mode reduces
$d_{x^2-y^2}$-superconductivity, because the  momentum
dependence of its coupling 
strength is peaked for large momentum transfer. On the other
hand the coupling strength of the buckling mode is peaked for small
momentum transfer and therefore the buckling mode supports
$d_{x^2-y^2}$-wave superconductivity in so far as the
$d_{x^2-y^2}$-pairing interaction is increased by the buckling mode.
The small wiggles in Fig.~\ref{fig_9} are caused by 
wiggles in the momentum dependence of the free susceptibility
$\chi^0_{\bf q}$, which determines the screening properties of the
CuO$_2$-planes (see appendix). In the case of the apex oxygen vibrations,
the momentum dependence of the coupling 
is different for the two apex ions in the primitive
cell: the closer apex oxygen causes a more or less momentum independent
coupling, whereas the coupling of the further apex oxygen is stronger
for small momentum transfer similar to the buckling mode like
vibrations. In respect to vibrations of the other ions, which are not shown in
Fig.~\ref{fig_9},  we find  that the coupling of vibrations in
z-direction is generally peaked for small momentum transfer. The
vibration of the oxygen ions is the most dominant
contribution to the total coupling strength, whereas the contribution of
yttrium and barium is very small due to screening in
the CuO$_2$-planes. Altogether all vibrations in z-direction,
especially the buckling modes, increase the nearest
neighbor interaction $V_{10}$. Only the breathing mode and the vibration
of the chain oxygen O(IV) along the $CuO$-bonding axis reduce
$V_{10}$. The other modes are of no relevance with respect to
$V_{10}$.

Note that within our approach the coupling strength of the buckling
mode is about twice as large as the coupling strength of the breathing
mode. The situation is opposite, if the same
calculations are performed with the original Zeyher-model. The reason
for this is to be found in the different treatment of screening. In
Zeyher's model the screening planes are smeared over the whole
elementary cell, whereas in our approach the exact position of the
$CuO_2$-planes with respect to the vibrating ion and the
Cooper-pairing hole is considered. It is evident that 

\noindent
the screening of the breathing mode is better, if one accounts for the fact that
both the vibrating oxygen ion and the Cooper pair forming hole are
located within the same plane. On the other hand the opposite is true for
the buckling mode. Because the vibration of the oxygen-
ion in z-direction is perpendicular to the $CuO_2$-plane, it
cannot be screened within the same plane, but only by charges induced
in more distant planes. Therefore, the buckling mode is
less screened, if one takes the relative location of the vibrating 
oxygen-ion, the Cooper-pairing
hole, and the screening plane  explicitly into account.
Additionally, it has to be pointed out that in our approach (as in the
Zeyher-model) only the electrostatic part of the electron-phonon coupling is considered,
whereas all contributions arising from changes in the overlap of
neighboring electronic wave functions have been neglected. It is to be
expected that these neglected effects are strongest for the breathing mode.
This means that the coupling strength of
the breathing mode could be somewhat underestimated in our
approach. In this respect it is interesting to
note that recently the electron-phonon coupling was calculated by
O.~K.~Andersen {\em et al.}~\cite{ASJ96} and  O.~Jepsen {\em et al.}~\cite{JAD97} within a tight-binding model
derived from LDA linear-response calculations. They
find a total electron-phonon coupling strength of $\lambda_s \approx 0.4$, which is 
similar to our findings, and a medium $d_{x^2-y^2}$-wave-coupling
strenth of $\lambda_d \approx 0.3$, which is even much larger than our
value. Most interesting, they also find a very dominant coupling of
the buckling mode. Although their approach is complementary to
ours, their results are quite similar, which supports our results
concerning the relative coupling strengths of the breathing and the
buckling mode. Anyway we  expect that a
more dominant breathing mode would reduce $V_{10}$ somewhat, leaving the other 
results qualitatively unchanged. 

\begin{figure}
\centerline{\epsfig{file=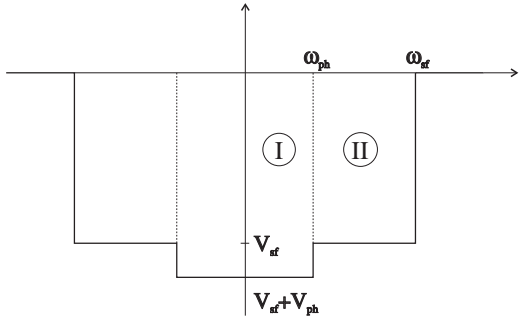,width=7cm,bb=110 480 460 730}} 
\caption{Modelling of the spin flucutuation interaction and the
  electron-phonon interaction by frequency independent coupling
  constants. Here, $V_{sf}=\lambda^d_{sf}$ and $V_{ph}=\lambda^d_{ph}$
  denote the $d_{x^2-y^2}$-type contribution of the corresponding
  interactions, respectively with cut off $\omega_{sf}$ and $\omega_{ph}$.}
\label{fig_10}
\end{figure}

\vspace{-.3cm}
Our numerical results, obtained within a strong coupling Eliashberg theory,
show that the electron-phonon interaction reduces the critical
temperature,  even though it increases the d-wave pairing interaction
and the isotope effect is positive. 
In order to investigate these aspects in 
a qualitative, but more transparent
way, we use a simple weak coupling  model, where
the pairing interactions caused by spin fluctuations and phonons are
approximated by frequency independent coupling constants $\lambda^d_{sf}$ and
$\lambda^d_{ph}$ with different cut-offs $\omega_{sf}$ and
$\omega_{ph}$ (Fig.~\ref{fig_10}). Introduction of the renormalization constants
$Z_{sf}=1+\lambda_{sf}$  
and $Z_{sf-ph}=1+\lambda_{sf}+\lambda_{ph}$ accounts for the reduction
of the spectral weight of the Cooper pair forming holes by scattering
effects due to spin fluctuations and phonons. For the renormalization factors the same
cut-off is used as for the pairing interaction. The coupling constants
$\lambda$ are defined in Eq.~\ref{eq:lambda}. The coupling constant
$\lambda^d$ characterize the $d_{x^2-y^2}$-pairing interaction:
\begin{equation}
\lambda^d= \frac{\sum_{\bf kk'} \delta(\epsilon_{\bf k})
     \delta(\epsilon_{\bf k'}) ReV({\bf k,k'},0)
     \psi_{\bf k}^d \psi_{\bf k'}^d }
     {\sum_{\bf kk'} \delta(\epsilon_{\bf k})
     \delta(\epsilon_{\bf k'}) \psi_{\bf k}^d \psi_{\bf k'}^d}\, 
\end{equation}
with $ \psi^d_{\bf k} = \cos k_x -\cos k_y $.
 These approximations result in the following form of the gap-function
 $\Delta_{\bf k}=\Delta_0 \psi^d_{\bf k} $
with $\Delta_0= \Delta_1$ if  $ |\epsilon_{\bf k}| < \omega_{ph}$
and $\Delta_0= \Delta_2$ if  $\omega_{ph} < |\epsilon_{\bf k}| 
                                                 < \omega_{sf}$, respectively.
$\Delta_1,\Delta_2$ are obtained by solving the   gap
equations: 
\begin{eqnarray}
\Delta_1 &=& \frac{\lambda^d_{sf}+\lambda^d_{ph}}{{\tt Z}^2_{sf-ph}}
    \int \limits_{0}^{\omega_{ph}} d\epsilon
           \frac{1-2f(\epsilon_1)}{\epsilon_1} \Delta_1 \nonumber \\ & &
  \,+\, \frac{\lambda^d_{sf}}{{\tt Z}^2_{sf}} 
        \int \limits_{\omega_{ph}}^{\omega_{sf}} d\epsilon
           \frac{1-2f(\epsilon_2)}{\epsilon_2} \Delta_2
 \end{eqnarray} 
and 
   \begin{eqnarray}              
\Delta_2 &=& \quad\enspace \frac{\lambda^d_{sf}}{{\tt Z}^2_{sf-ph}} \quad\,
    \int \limits_{0}^{\omega_{ph}} d\epsilon
           \frac{1-2f(\epsilon_1)}{\epsilon_1} \Delta_1 \nonumber \\ & &
  \,+ \enspace\,\, \frac{\lambda^d_{sf}}{{\tt Z}^2_{sf}} \enspace\,
        \int \limits_{\omega_{ph}}^{\omega_{sf}} d\epsilon
           \frac{1-2f(\epsilon_2)}{\epsilon_2} \Delta_2  \enspace.         
\label{eq:gap_eq}
\end{eqnarray} 

\begin{figure}
\centerline{\epsfig{file=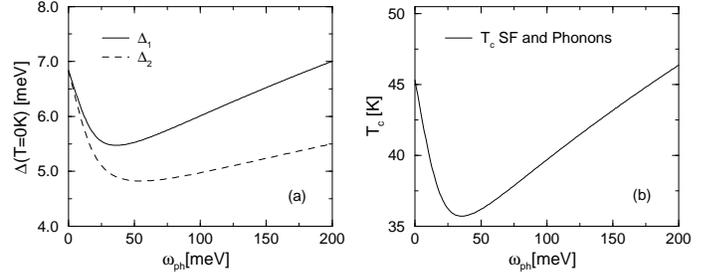,width=10cm,bb=0 0 570 250}} 
\caption{Results for $\Delta$ and $T_c$ using the simplified
  interaction model taking $\omega_{sf}=200 meV$,
  $\lambda^d_{sf}=1.1$, $\lambda_{sf}={3.2}$, $\lambda^d_{ph}=0.3$,
  $\lambda_{ph}=0.6$.} 
\label{fig_11}
\end{figure}

\noindent
with
\begin{equation}
\epsilon_1=\sqrt{\frac{\epsilon^2}{{\tt Z}^2_{sf-ph}}+\Delta_1^2}
\quad , \quad
\epsilon_2=\sqrt{\frac{\epsilon^2}{{\tt Z}^2_{sf}}+\Delta_2^2}
\end{equation}

In the following we will study the influence of the size 
of the phonon cut-off $\omega_{ph}$ on the superconducting properties by means 
of this simple model. In Fig.~\ref{fig_11}(a) the 
gap-functions $\Delta_1$ and $\Delta_2$ corresponding to the regions I
($|\epsilon_{\bf k}| < \omega_{ph}$)
and II ($ \omega_{ph} < |\epsilon_{\bf k}| < \omega_{sf}$)
at $T=0K$ are plotted and in ~\ref{fig_11}(b) the
corresponding critical temperature is shown. 
The coupling constants $\lambda_{sf}$, $\lambda^d_{sf}$,
$\lambda_{ph}$, $\lambda^d_{ph}$ are chosen in agreement with the
previous results. Only $\lambda^d_{ph}$ is chosen slightly
larger for illustrational purposes. Since the spin fluctuation interaction is decreasing
 slowly  over a large frequency range, as can be seen by inspection of
Fig.~\ref{fig_1}, it is hard to estimate the spin fluctuation
cut-off $\omega_{sf}$. In this calculation $\omega_{sf}=200meV$ was
chosen as a cut-off. Two things can be observed in
Fig.~\ref{fig_11}: First, the gap-function $\Delta_1$ (region I)
is generally larger than the gap-function $\Delta_2$ (region II)
due an increased pairing interaction caused by
phonons. Furthermore the gap-functions $\Delta_1$ and $\Delta_2$ and
the critical temperature $T_c$ show a peculiar
$\omega_{ph}$-dependence: for small $\omega_{ph}$ they are decreasing
as a function of $\omega_{ph}$ and after a certain critical cut-off
frequency they are increasing again. 
Examination of gap-equation Eq.~(\ref{eq:gap_eq}) shows
that there are two contributions,
which determine the size of the gap-function $\Delta_1$: the interaction
of charge carriers(I) among themselves and the interaction of charge 
carriers(I) with charge carriers(II). In an analogous manner there
are two contributions, which determine the size of the gap-function
$\Delta_2$. 
Only one of these (four) contributions, i.e both electrons in region
I, can profit from the phononic part of the pairing interaction. On
the other hand two of these (four) contributions, each of which
result from the interaction of the respective electrons with electrons of region I, are
reduced, because the quasi particle spectral weight in region I is
diminished by the additional scattering on phonons. Altogether only the
gap-function $\Delta_1$ can profit from the phononic contribution to the
pairing

\begin{figure}
\centerline{\epsfig{file=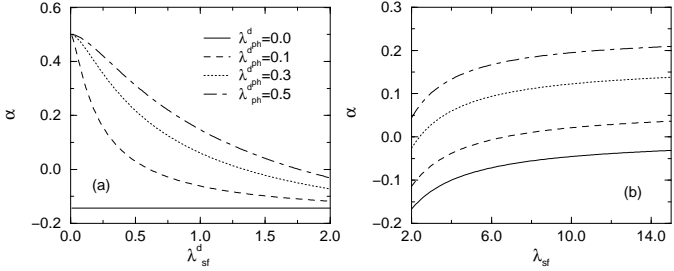,width=9.2cm,bb=15 0 570 250}}
\caption{Dependence of the isotope exponent $\alpha$ on the spin fluctuation
  $d_{x^2-y^2}$-coupling strength and the total spin fluctuation
  coupling strength, respectively. These results were obtained with
  the simple weak coupling
  model with $\omega_{sf}=200meV,\omega_{ph}=70meV$ and additionally
  (a) $\lambda_{sf}=2.5$ and (b) $\lambda^d_{sf}=1.3$, respectively.
  The total electron-phonon coupling strength was taken as $\lambda_{ph}=1$,
  but the $d_{x^2-y^2}$-electron-phonon coupling strength was varied
  as indicated in the figure.}
\label{fig_12}
\end{figure}

\noindent
interaction, whereas both gap-functions $\Delta_1$ and
$\Delta_2$ notice the destructive effect due to scattering on phonons.
This is the reason why for small $\omega_{ph}$ both of the
gap-functions $\Delta_1$ and $\Delta_2$ are decreasing as a function of
$\omega_{ph}$. The larger region I the more electrons
are scattered by phonons. Only when region I has become large enough
and $\omega_{ph}$ exceeds a critical cut-off frequency, the
positive effect of increased pairing interaction between Cooper
pairs in region I outweighs the negative effect of additional scattering
and thus both gap-functions $\Delta_1$ and $\Delta_2$ start increasing as a
function of $\omega_{ph}$ for larger frequencies.

The $\omega_{ph}$-dependence of $T_c$ shown in
Fig.~\ref{fig_11}(b) displays an analogous behavior.
For very small $\omega_{ph}$ the critical temperature is decreasing,
and after passing through a minimum, $T_c$ is increasing again yielding a
positive isotope effect. The critical temperature
itself will, for realistic phonon frequencies $\omega_{ph}$, be
smaller than the critical temperature of only
spin fluctuation induced superconductivity.

From  Fig.~\ref{fig_11}(b) we can directly extract the isotope
exponent. For very small phonon
frequencies the critical temperature is decreasing for increasing
$\omega_{ph}$. Hence, the isotope exponent $\alpha$ is negative in
this region. For larger $\omega_{ph}$, after passing through a minimum,
$T_c$ is increasing with $\omega_{ph}$. This causes a positive isotope
exponent. 
The magnitude of $\alpha$ depends on the
relative size of the coupling strengths of the electron-phonon and
spin fluctuation interaction and the corresponding scattering rates. In
Fig.~\ref{fig_12} the isotope exponent vs $\lambda^d_{sf}$ and
$\lambda_{sf}$ is shown for different values of $\lambda^d_{ph}$. For
decreasing spin fluctuation contribution $\lambda^d_{sf}$ to the pairing interaction
 the isotope exponent is increasing strongly. This is obvious, 
  because the electron-phonon coupling becomes the origin for
superconductivity, a situation which seems unlikely for cuprates 
due to the large phononic coupling constants necessary in this case.

The isotope exponent of the present work  is always small in magnitude
and positive. For the system YBa$_2$Cu$_3$O$_{7}$, under
consideration, this is in qualitative agreement with the experimental
observation. Whether the increase of $\alpha$ for non-optimal doping
concentration is solely due to the decreasing strength of the leading
electronic pairing interaction  or of more subtile nature is beyond
the scope of the present paper.

\section{Summary}

In summary, we find that spin fluctuation induced
$d_{x^2-y^2}$-superconductivity is robust against the existence of
electron-phonon coupling. The superconducting
transition temperature is reduced by the interaction with
phonons due to an enhancement of the electron scattering rate.
Furthermore, we find that the momentum dependence of the
electron-phonon interaction increases the
$d_{x^2-y^2}$-pairing interaction slightly. This is important
for getting a positive isotope effect. Our results indicate that 
the isotope effect is small for optimal doping. If
one assumes that the spin-fluctuation coupling strength varies with
the doping concentration, this would imply a strong doping
dependence of the isotope effect.

A major approximation of the present work is the neglect of the effect of
strong electronic correlation on the electron-phonon coupling
constants. As shown by Lichtenstein and Kuli\'c~\cite{LK95}
strong electronic correlation cause a suppression of charge
fluctuations at small distances. Therefore, 
we expect $g_{\lambda}({\bf k})$ to be reduced for zone
boundary phonons. This would decrease $\lambda^s_{ph}$ and thus
the large scattering rate due to phonons. For the same
reason, we expect $\lambda^d_{ph}$ to be slightly enhanced by
electronic correlations. 

Our results demonstrate that the existance and sign   of the
isotope effect   can be understood within a spin fluctuation induced
d-wave state. Due to the d-wave nature of the superconducting state, it  is
particularly important to obtain a proper description of the momentum
dependence of the electron-phonon interaction.
It was shown that  out-of-plane vibrations are 
 essential   for the d-wave state and that one cannot 
focus solely on one particular phononic  mode.
 Finally, the strong
electronic correlations  inhibit the occurrence of a phonon
induced superconducting state with s-wave symmetry.
A phonon induced d-wave state can be excluded because of
the insufficient strength of this subdominant pairing interaction.

\section{Acknowledgement}
We are  grateful to M. Peter, D. Pines, M. Kuli\'c, B. Stojkovi\'c,
 and C. Thomsen for helpful
conversations on this and related topics.
 J. S. acknowledges the financial support of the Deutsche
 Forschungsgemeinschaft.

\begin{appendix}
\section{Derivation of the inverse dielectric function}

In this appendix the treatment of the local field corrections arising
from the position of the screening $CuO_2$-planes relatively to the
vibrating ion will be discussed.
If spatial inhomogenity   is taken into account, the dielectric function
becomes a matrix in the reciprocal lattice vectors ${\bf G,G'}$:
\begin{eqnarray}
\epsilon({\bf k+G,k+G'}) &=& \epsilon_{\infty} \delta_{\bf GG'} -
\nonumber \\ & &
               v({\bf k+G}) \tilde \chi({\bf k+G,k+G'}) \enspace .
\label{eq:Matrix_epsilon}
\end{eqnarray}
Here, $\tilde \chi$ is the irreducible polarization of the
density-density correlation function $\chi$, which characterizes the
screening properties of the system.
The inverse dielectric function is defined by:
\begin{equation}  
\sum_{\bf G''} \epsilon({\bf k+G,k+G''}) \epsilon^{-1}({\bf k+G'',k+G'})
                    = \delta_{\bf GG'}  \enspace .
\end{equation}
This can be expressed as:
\begin{eqnarray}
& &\epsilon^{-1}({\bf k+G,k+G'}) = \frac{1}{\epsilon_{\infty}}
                         \left(
      \delta_{\bf GG'}\right. \nonumber \\ 
\lefteqn{\left.  + v({\bf k+G}) \chi({\bf k+G,k+G'})  \right) }\, .
\end{eqnarray}
The susceptibility $\chi$ has to be determined by solving the
following integral equation:
\begin{eqnarray}
\label{eq:Dyson-chi}
& &\epsilon_{\infty} \chi({\bf k+G,k+G'}) =
    \tilde \chi({\bf k+G,k+G'}) +   \nonumber     \\
\lefteqn{ \sum_{\bf G''} \tilde \chi({\bf k+G,k+G''})  
           v({\bf k+G''}) \chi({\bf k+G'',k+G'}). } \nonumber \\ 
\end{eqnarray}
In the cuprates the most dominant screening is caused by mobile charges within the
CuO$_2$-planes. In YBa$_2$Cu$_3$O$_7$ there are two planes per unit
cell. Therefore, the susceptibility $\chi$ is a $2 \times 2$-matrix,
describing density-density correlations within the same plane and between 
both of the planes.
The polarization $\tilde \chi$ on the other hand reduces to a diagonal
matrix, because the particles are assumed to be confined to the planes. In real space the
polarization is given by:
\begin{equation}
\tilde \chi({\bf r_1,r_2})   
            = \tilde \chi^{2D} ({\bf x_1 - x_2}) \delta (z_1 - z_2)
              \sum_{j \beta} \delta (z_1 - z_{j \beta})  \enspace ,
\end{equation}
where $z_{j \beta}=jc+d_{\beta}$ are the z-coordinates of the
$CuO_2$-planes, with $j$ being the index for the primitive cell, $\beta$
the index for the $CuO_2$-plane within the primitive cell and c
the height of the primitive cell.
Fourier transformation yields:
\begin{equation}
\label{eq:chi-tilde}
\tilde \chi({\bf k+G},q+Q,q+Q') =
      \frac{1}{c} \tilde \chi^{2D}({\bf k+G}) 
      \sum_{\beta} e^{i(Q-Q')d_{\beta}}
      \enspace .
\end{equation}
Here, ${\bf k,G}$ refer to the in-plane wave vectors, whereas $q,Q$ represent
the wave vectors in z-direction. In RPA $\tilde \chi^{2D}$ reduces
to the noninteracting particle-hole bubble of the holes in the
$CuO_2$-planes. $\tilde \chi$ is diagonal with
respect to the in-plane wave vectors and the $Q$-dependence is
simple enough. This permits an analytic treatment of the
$Q$-summations. Therefore, Eq.~(\ref{eq:Dyson-chi}) can be inverted
easily and the inverse dielectric function can be obtained in this
way. 
There remains only the question of how to choose the z-coordinate
$d_{\beta}$ of the screening planes, which are not unambiguous  due to 
the internal buckling of the $CuO_2$-planes. In our calculations we
assumed that the screening plane should follow the
buckling of the $CuO_2$-plane.

\end{appendix}

%
%
%

\end{multicols}

\end{document}